# Hydration Peculiarities of Graphene Oxides with Multiple Oxidation Degrees

Antenor Neto[1], Vitaly V. Chaban,[2] and Eudes E. Fileti[3]

[1] Centro de Ciências Naturais e Humanas, Universidade Federal do ABC, 09210-170, Santo André, SP, Brazil.

[2] P.E.S., Vasilievsky Island, Saint Petersburg, Russian Federation.

[3] Instituto de Ciência e Tecnologia, Universidade Federal de São Paulo, 12231-280, São José dos Campos, SP, Brazil.

**Abstract.** Hydration properties of graphene oxide (GO) are essential for most of its potential applications. In this work, we employ atomistic molecular dynamics simulations to investigate seven GO compositions with different levels of oxygenation. Two atomic charge models for GO are compared: (1) $sp^2$ carbons are purely Lennard-Jones sites; (2) $sp^2$ carbon charges are consistent with the CHELPG scheme. Structural properties were found to depend insignificantly on the charge model, whereas thermodynamics appeared very sensitive. In particular, the simplified model provides systematically stronger GO/water coupling, as compared to the more accurate model. For all GO compositions, hydration free energies are in the range -5 to -45 kJ mol$^{-1}$ indicating that hydration is thermodynamically favorable even for modest oxidation degrees, thus differing drastically from the case of pristine graphene and graphite. The results and discussion presented hereby provide a physical background for modern applications of GO, e.g. in electrodes of supercapacitors and inhibitors in processes involving biological molecules.

**Introduction**

Graphene oxide (GO) has been considered a convenient substitute for graphene in important technological applications due to its remarkable electrical, mechanical and thermal properties.[1-10] The advantages of GO in relation to pure graphene are due to the drastic structural and electronic changes resulting from the functionalization of graphene with oxygen groups.[1-10]

It is well known that the detailed atomic structure of GO is very difficult to obtain since this material is non-stoichiometric, presenting a wide variety of compositions which depend inherently on the route of synthesis.[5, 11-14] In addition, GO is strongly hydrophilic, hygroscopic and thermally unstable at intermediate temperatures (60-80$^{o}$C).[5, 11-14] All this makes the controlled synthesis of this material very difficult and the exact composition for each degree of oxidation uncertain and strongly dependent on the countless possible combinations for oxygen coverage. These combinations are related to the concentration of oxygen on the surface, defined by the O/C ratio, the epoxy/hydroxyl ratio, the edge functional groups and, lastly, the uniformity and regularity of the distribution of the groups on the basal plane.[5, 14, 15]

Unlike pristine graphene, GO shows a better solubility in water due to functional epoxy and hydroxyl groups that functionalize the basal plane. It is generally expected that the interesting properties of GO, in particular its water solubility, will depend significantly on the degree of graphene oxidation. Most works on graphene oxide applications either treat only a single composition for GO or when varying their composition only discuss inherent structural or electronic properties. However, it is possible to find some works where applied properties of GO are considered according to the degree of oxidation of the GO. Supercapacitors based on GO/ionic liquids have been investigated by molecular dynamics simulations.[3, 16] It has been observed that both the quantity and the type of oxygen-containing functional groups on the graphene surface can influence supercapacitor performance due to specific interactions of ions with GO functional groups and resulting local ion concentration changes near the GO electrode.[3] Mu and colleagues investigated the dependence of the thermal conductivity of GO with coverage rate of oxygen groups and observed that the thermal conductivity decreases as the concentration of oxygen increases.[17] In that work a conductivity variation was observed in a range of five orders of magnitude, showing an enormous tunability which is highly desirable for the production of efficient thermoelectric materials.[17] Water fast slip flow and wettability properties on GO surface were investigated by molecular dynamics in function of oxidation degree.[18, 19] A significant flow rate enhancement, by more than two orders of magnitude, relative to pristine graphene, was observed in nanoconfinement in GO.[18] In addition, the hydrophilicity promoted by oxygen-containing groups on graphene leads to a decrease in contact angle with increasing concentration.[19]

Solubility in water is certainly one of the most relevant requirements for any candidate system for biomedical applications. For GO, it is even more relevant since this material presents unique geometry that favors hydrophilic interaction with water and/or biomaterials on both sides, which facilitates access for covalent and non-covalent functionalization in addition to efficient loading of molecules, from small organic ones to biomacromolecules.[4, 20-24] In this work we employ atomistic molecular dynamics simulations to investigate how the hydrophilic character of GO is affected by the increase of the concentration of oxygen groups on its surface. For this we investigate in detail its structure, interaction with water as well as the thermodynamic of the hydration through free energy calculations.

**Simulation Details**

It is widely accepted that graphene oxide basically consists of epoxy and hydroxyls groups adsorbed on a basal carbon plane.[5] The ratio between the number of carbon and oxygen atoms in the sample defines the oxygen coverage ratio, R; that is $R = 100 \times cO/cC$. Although from a theoretical point of view the proposed models assert that such functional groups distribute in an orderly manner,[5, 14, 15] more recent measurements indicate that amorphous models of GO are the ones that best describe the experimental results.[5, 14, 15]

In this work, the properties of hydration of the GO are investigated as a function of the oxygen coverage ratio assuming a ratio between the number of hydroxy and epoxy groups to be fixed; cOH/cO = 2. For this we consider seven oxygen coverage ratios ranging from 10 to 70% (see Figure 1) which cover all experimentally observed compositions.

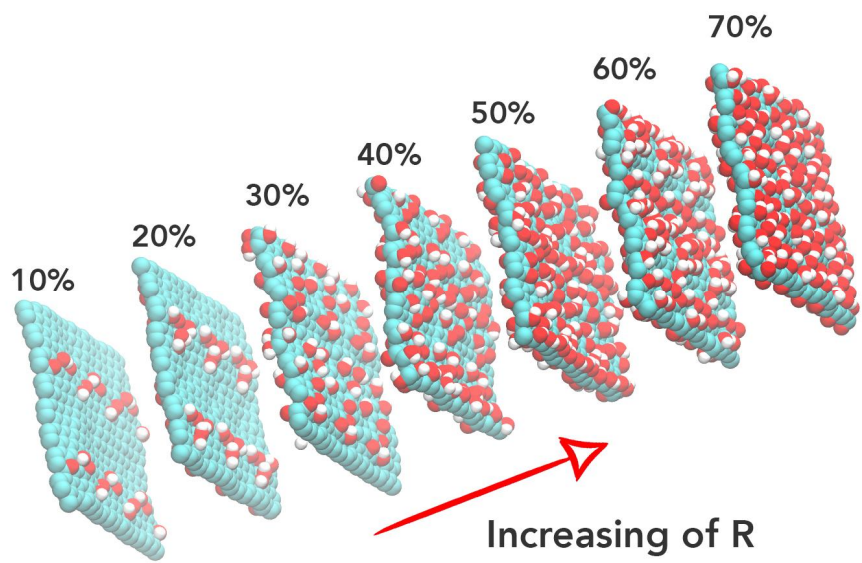

**Figure 1**. Oxygen concentration ratios (R) for the graphene oxide models employed here. This ratio is defined by the oxygen coverage defined by $R = 100 \times cO/cC$, where $cO$ and $cC$ are the numbers of oxygen and carbon atoms in the sample, respectively.

The configurations for the GOs at each of these concentration ratio were taken from the work of Chen et al.[5] They generated these configurations following rules to obtain amorphous models, namely: *i*) in each carbon atom it is possible to adsorb only one functional group; *ii*) for each side of the graphene plane, paired hydroxyls are added on adjacent carbon atoms; *iii*) due to the steric effects, no more than four carbon atoms attached to hydroxyl groups or five carbons attached to epoxy groups in a single six-carbon ring and *iv*) in order to reduce the strain effects, the number of groups in both sides of the basal plane should be approximately the same. The structures generated following these rules were relaxed using periodic first-principles DFT computations with plane-waves and pseudopotentials.[5] From these structures were obtained an infinite and finite model for the GO. The infinite model was obtained by replicating the original cell from Chen[5] in a 3x2 supercell and placed in a computational cell with appropriate distances x and y to ensure chemically correct connectivity of the edges. The ketone and carboxylic groups, as well as the defects in hole shapes, typically found on the edges of finite samples were not considered. The finite model was generated by a smaller number of replicates (2x1) of the original cell. As our main interest is to describe how the hydrating properties of GOs vary with the oxygen coverage ratio, the edges of the finite sheet were saturated by applying a united site model to the edge carbons. This allowed us to analyze only the effect of the surface groups (epoxy and hydroxyls) on the hydration process avoiding the drastic effects of the highly polar groups that usually functionalize the GO edges. Figure 2 presents both GO models in their respective computational cells.

Classical molecular dynamics simulations of graphene oxide have been usually performed employing simplified models where the $sp^2$ carbon in GO is treated as uncharged Lennard-Jones spheres.[3, 9, 16, 18, 20, 22, 25-30] However the adsorption of epoxy or hydroxyl groups on the surface of the pristine graphene leads to a rearrangement of charges in the carbon planar structure, drastically altering the electrostatic character of the involved sites. In order to take into account such charge redistribution in our model we calculate the partial electric charges for each system investigated here using quantum mechanical calculations. Such electric charges were obtained at B3LYP/6-31G(d,p) theoretical level using CHELPG[31] scheme on the relaxed structures for the 2x1 saturated system and then manually transferred to larger systems 3x2 (infinite sheet). We take care to consider only the charges of the atoms of the central region of the 2x1 structure before replicating the charges, such that we practically eliminate the edge effects on the charge edge-atoms due the finite GO structure. The electronic structure computations were performed in Gaussian 09, revision D.[32] The interaction model used to describe GO were treated by a CHARMM36 based force field.[33] Spring constants as well as sigma and epsilon Lennard-Jones parameters were taken directly from CHARMM36. The water molecules was modeled using the TIP3P model.[34] To evaluate the effect of the induced charges on the $sp^2$ carbon atoms (neighboring to the oxygenated sites) we also performed all the calculations employing a simplified model. In this model, we treated all $sp^2$ carbon atoms as uncharged LJ sites while the partial charges on the other atoms were taken as an average value over all charges (see supporting information) obtained for the seven different systems using CHELPG scheme. Thus, the charges for the O(epoxy), O(hydroxyl), H(hydroxyl) and C($sp^2$) atoms were respectively -0.36e, -0.70e, +0.40e and +0.30e. These average values are similar to those determined by Stauffer, by the DFT charge scaling, which was performed to take into account the polarization effect of an aqueous environment.[23]

Two different simulation series were performed to determine the structure, energetic and thermodynamics of the hydration of GO. In the first series of simulations, an infinite GO sheet was immersed in a periodic computational cell containing about 1500 water molecules. The simulations were performed in the isobaric-isothermal ensemble (constant pressure and temperature, NpT) under conditions T = 298 K and p = 1 atm. The simulation cells were initially subject to energy minimization aimed at the removal of high-energy contacts. The production stage was performed during 22 ns, using a time step of 2 fs. The first two nanoseconds of the trajectory were discarded from the analyses, being considered an equilibration stage. Configurations of the systems were saved every 2 ps totaling to 10000 frames for statistical analysis. The molecular representation a representative configuration of a simulation cell is shown in Figure 2.

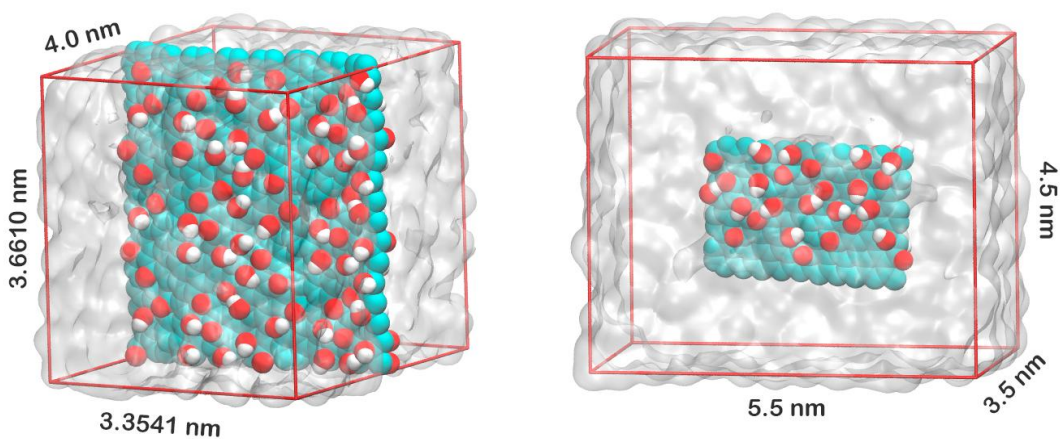

**Figure 2.** At left, a representative configuration of the simulation cell used to calculate the energy and distribution profiles of an infinite graphene oxide sheet. At right, the model used to calculate the hydration free energy of the finite graphene oxide model. The presented model is for R = 40%. The dimensions of each box are shown. Note that the x and y dimensions of the box with the infinite sheet are accurate to ensure the connectivity of the sheet edges through the periodic boundary conditions.

The electrostatic interactions beyond 1.2 nm were accounted for by Particle-Mesh-Ewald (PME) method.[35] The Lennard-Jones interactions were smoothly brought down to zero from 1.1 to 1.2 nm using the classical shifted force technique. The constant temperature was maintained by the velocity rescaling thermostat[36] (with a time constant of 1 ps), which provides a correct velocity distribution for a statistical mechanical ensemble. The constant pressure of 1 atm was maintained by Parrinello-Rahman barostat[37] with a time constant of 2.0 ps and a compressibility constant of $4.5 \times 10^{-5}$ bar$^{-1}$.

The second series of simulations aimed at obtaining the hydration free energy of GO. For this we use the Bennett Acceptance Ratio (BAR) method,[38] a slow-growth procedure that allows the gradual decoupling of the GO from its equilibrium aqueous environment by the calculation of the $\langle \frac{dH(\lambda)}{d\lambda} \rangle$ where $H$ is the parameterized empirical Hamiltonian and $\lambda$ is a coupling parameter. $\lambda = 1$ corresponds to the fully solvated GO, whereas $\lambda = 0$ corresponds to the non-interacting solute and solvent. This procedure was divided into 31 $\lambda$-states. In the first 10 states, the GO-water electrostatic interactions were deactivated using an increment of $\Delta\lambda = 0.1$. Then the van der Waals interactions have been turn off in the last 21 states using $\Delta\lambda = 0.05$. For the purpose of avoiding singularities, we have used the soft-core interactions for the LJ interactions:[39]

$$V_{SC}(r) = (1 - \lambda)V([\alpha\sigma^6\lambda^p + r^6]^{1/6})$$

where $V_{SC}(r)$ is the normal hard-core pair potential and $\sigma$ is the LJ size parameter of the atom pair. The parameters for the soft-core were $\alpha = 0.5$, $p = 1.0$, and $\sigma = 0.3$.

For every $\lambda$, the systems were equilibrated during 1 ns with production stage of 5 ns using the same simulation parameters as in the simulations from first series. The only exception is that, for the sake of proper sampling, stochastic dynamics was used instead of the conventional dynamics. For the Langevin thermostat we have used a friction coefficient of 1 ps$^{-1}$. All trajectories were propagated using the GROMACS 2016.2 simulation software.[40]

**Discussion**

The mass distribution profiles show how the water molecules distribute on the surface of GO, see Figure 3. It can be observed that for smaller concentration ratios (R = 10-40%) the waters are ordered with two clear structuring peaks before reaching the bulk. In these cases, in addition to interaction with the polar groups from graphene epoxy, the water molecules also interact directly with the carbon atoms of the free regions. This interaction tends to be hydrophobic thus allowing a greater structuring of the water on the GO surface. For higher concentration ratios (R = 50-70%) this behavior is still expected, however the greater number of hydrophilic sites for interaction between the water molecules with the GO surface reduces and displaces both the first and the second hydration peaks of the mass distribution.

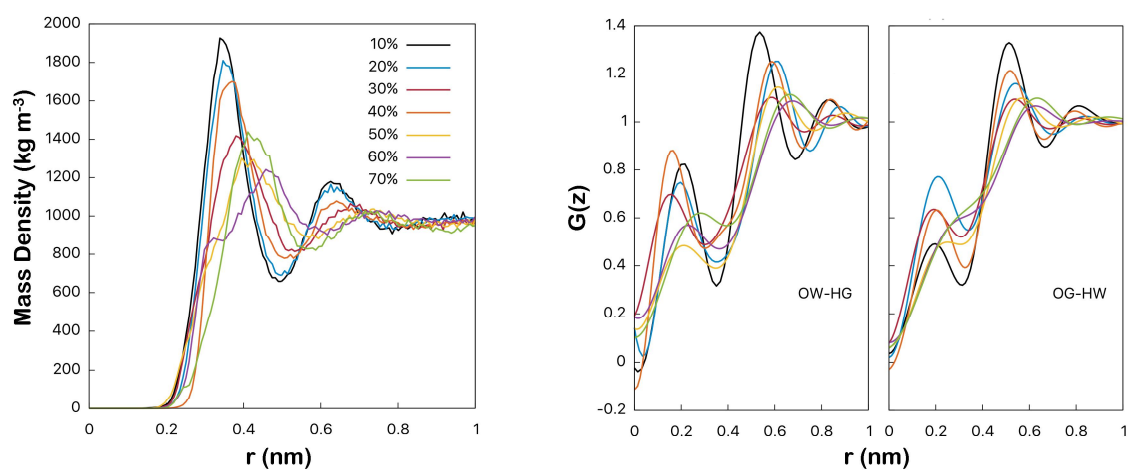

**Figure 3**. At left, mass density profiles of the water near at the grafene oxide surface (in kg m$^{-3}$) for all concentration ratios R. At right, O⋯H pairs distribution as a function of the z-distance from the graphene oxide surface. OW-HG represents the pair formed by water oxygen and GO-hydroxyl hydrogen and OG-HW the pair formed by GO-oxygen and water hydrogen.

It is known that GO and water have a strong hydrophilic interaction governed by the mutual formation of hydrogen bonds (HBs). The distribution of O⋯H pairs as a function of the z-distance from the surface, G(z), is also given in Figure 3. As can be seen there are two clear peaks, the first one related to the O⋯H distance in a hydrogen bond and the second related to the position of the hydrogen atom of the H-bonded neighbor. The positions of both peaks, in 0.20 nm and 0.32 nm, are very close to that of the bulk water allowing to conclude that the structure of the water adsorbed on the GO surface is similar to bulk liquid water. Although the positions of the G(z) peaks are similar to those found in pure water, the height of the peaks are not, in fact they are significantly smaller than the corresponding ones for the bulk water. This is related to the lower number of HBs that water formed with GO in relation to the number of HBs formed in bulk. This number of HBs depends not only on the R but also on the number of epoxy and hydroxyl groups, which have been chosen at random for our systems. Thus it is not possible to obtain a clear relation between the concentration of oxygen and the height of the G(z) peak. However, in general, we can observe that at low oxygen concentration ratios (R = 10-40%) the peaks are higher and in high oxygen concentration ratios (R = 50-60%) the peaks are smaller. The total number of HBs formed between water and GO can be obtained by adding the integral of the first peak in both distribution, $G(z)_{OW-HG}$ and $G(z)_{OG-HW}$. This value will be analyzed later. When obtained from the simulations using the simplified model for the GO's, both mass density profiles and G(z) showed the same qualitative behavior and only slight quantitative differences, therefore they will be omitted here.

To get a better understanding of the energetics of the GO-water interaction, we calculate the total interaction energy per area unit, see Figure 5. In these plots we decomposed the interaction energy into its electrostatic (Coulomb) and van der Waals contributions. In addition, we also decomposed the energy in terms of the two GO sites (C for graphene structure and O/OH for hydrophilic adsorbed groups). In this way, we calculated the contributions of the carbon structure separately from the contributions of the polar sites.

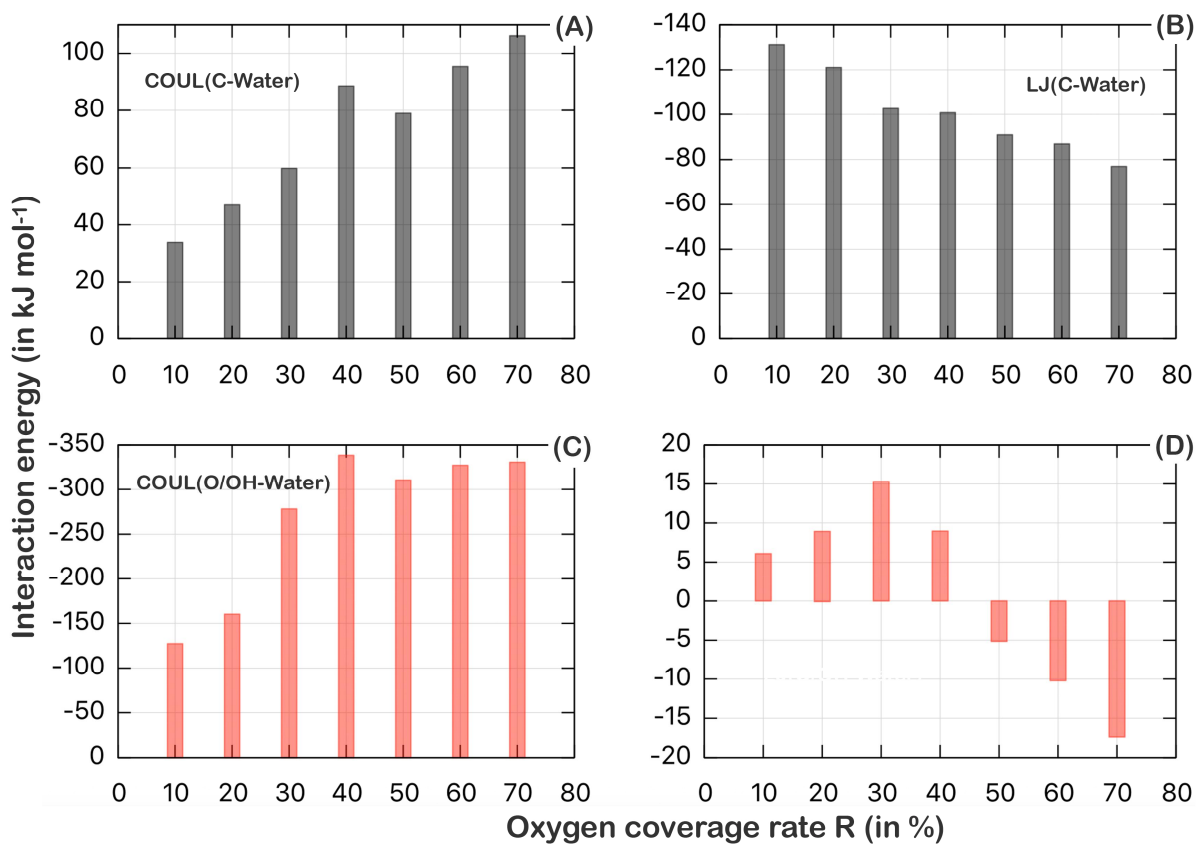

**Figure 4**. Decomposition of the total interaction per area (in kJ mol$^{-1}$) between water and GO in terms of GO sites (carbon in black and O/OH groups in red) and also in terms of the electrostatic components (Coul) and van der Waals (LJ).

Overall, we observed that the greatest contribution to the interaction energy between water and GO comes from the electrostatic interaction of the hydrophilic groups (chart C) and the van der Waals interaction of the carbon structure (chart B) for any R ratios. As can be seen, these two contributions together are greater than the total energy. This is possible because there are repulsive contributions to interaction energy coming mainly from the carbon structure. The models employed here have a clear advantage. When considering the redistribution of charges on graphene after oxidation, the carbon surface has become slightly charged so that its contribution to interaction energy is positive. The increase of this repulsive contribution (chart A) occurs in a linear way with the concentration while that the corresponding van der Waals energy linearly decreases (chart B). We also observed that for reduced graphene oxide (low values of R) the small contribution of van der Waals from O/OH groups is also positive (Chart D). This is because for reduced graphene oxide the water molecules have a greater proximity to the adsorbed groups, lying within the repulsive region of the Lennard Jones model.

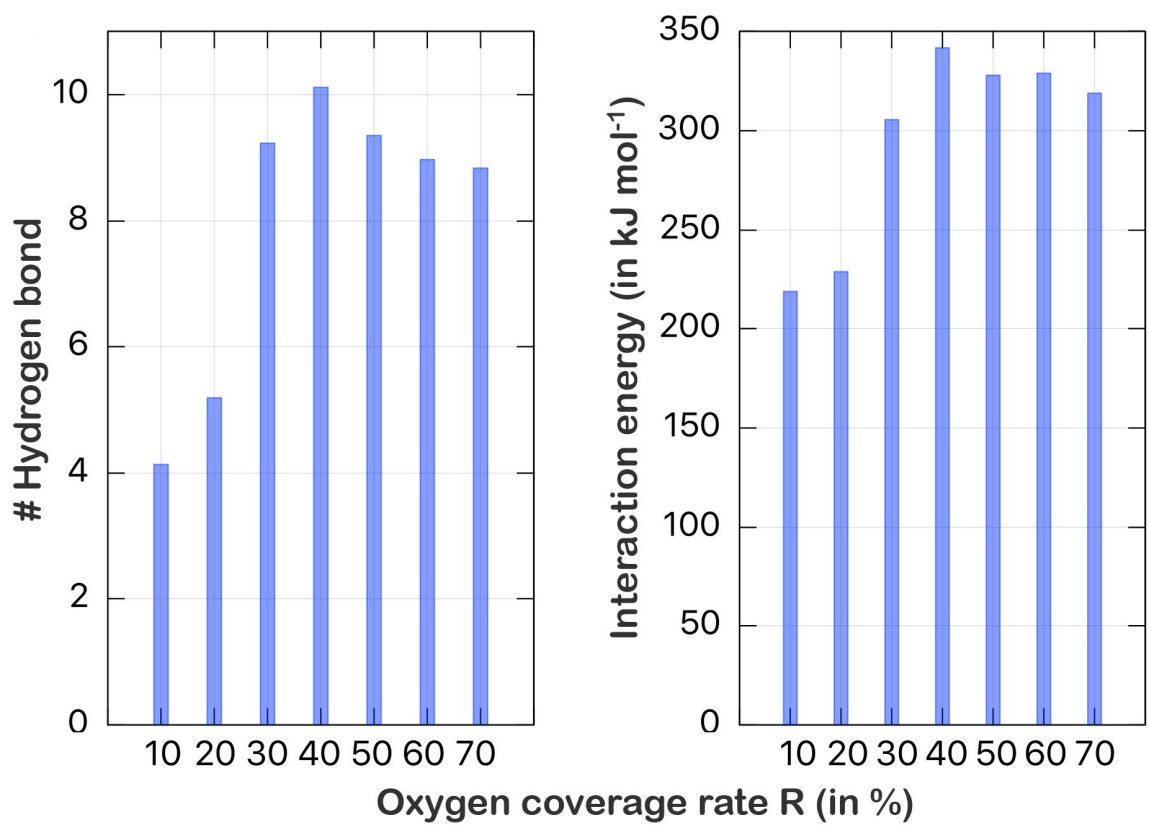

**Figure 5**. Number of hydrogen bonds and total energy of interaction (in kJ mol$^{-1}$), per unit area, between water and graphene oxide.

The total interaction energy (sum of the bars of the charts A-D from Figure 5), normalized per unit area, between GO and the aqueous medium, are shown in Figure 6, along with the average number of formed hydrogen bonds. A linear increase of both, number of HBs and total energy of interaction could be expected as with other homologous systems.[41, 42] However, here we observe that these averages reach a maximum in R = 40% and stop growing, even presenting a slight decrease. This is possibly due to steric effects that prevent the formation of a greater number of hydrogen bonds between the O and OH groups of the GO with the water molecules. The correlation between the number of HBs and the total energy of interaction is clear, with this varying according to the amount of HBs formed. For the concentration R = 40% we observed an average number of 10.1 hydrogen bonds with a total energy of interaction of -342 kJ mol$^{-1}$, both per unit area. Comparison between the two models (CHELPG and average charges) reveals that while LJ interactions, as expected, present virtually no change, the electrostatic interactions present changes that vary from 1 to 13% depending on the analyzed system. For the largest variation (R = 20%), we observed that the average-charge model overestimates the energy of water interaction by about 36 kJ mol$^{-1}$ per unit area. The total number of hydrogen bonds is also significantly affected, ranging from 2% to 12%. For example, for GO with R = 30%, using the CHELPG model we found 9.1 hydrogen bonds per unit area while for the average-charge model we found 10.4 bonds per unit area.

A good basis for the understanding of the interactions between water and GO can be obtained, as done in the previous section, by the analysis of the interaction pairwise energies. However we must be aware that this analysis is still somewhat superficial, taking into account only the potential energies. For a precise thermodynamic analysis, we must take into account not only the enthalpic aspects, related to the interaction energy between the molecular species, but also to the entropic aspects, related to rupture of the hydrogen bond network in the bulk water due to the presence of GO. The thermodynamic potential that takes into consideration these aspects is the hydration free energy, $\Delta G_{hyd}$.

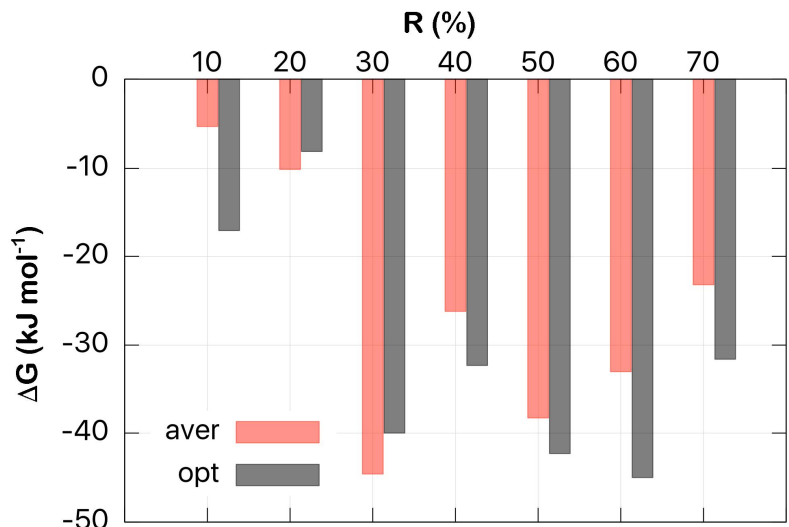

**Figure 6**. Hydration free energy per unit area ($\Delta G_{hyd}$, in kJ mol$^{-1}$) in function of oxygen concentration R. Black and red bars stand for CHELPG (opt) and average (aver) charge models.

Figure 6 shows the hydration free energy (per area unit) for each analyzed system. In general, the $\Delta G_{hyd}$ values are in the range of -5 to -45 kJ mol$^{-1}$ per unit area indicating that hydration is a thermodynamically favorable process for all investigated systems. However, we observe that there is no clear dependence between $\Delta G_{hyd}$ value and the oxygen concentration ratio, R. For the CHELPG values (black bars), the highest $\Delta G_{hyd}$ value occurs for R = 60% while the lowest occurs for R = 20%. On the other hand, for the values obtained with average charges the highest value for $\Delta G_{hyd}$ occurs for R = 30%. The absence of a well-defined tendency for $\Delta G_{hyd}$ of graphene oxide differs drastically from the behavior observed for homologous series of other polyhydroxylated systems, such as polyols and fullerenols for which $\Delta G_{hyd}$ vary practically linearly with the number of hydroxyl groups.[41, 42] A possible explanation for this lies in the GO topology. Reduced GO (R < 20%) consists of a plane of carbons, with islands of polar groups separated by hydrophobic regions. This peculiar structure confers to GO amphiphilic character and depending on its dimensions and the edge saturations it can be a hydrophilic or hydrophobic system. This characteristic seems to particularly affect $\Delta G_{hyd}$ values for GO with R = 10% and 20% where higher hydrophobic regions are exposed to the aqueous environment.

Despite the lack of well-defined pattern, our results suggest that GO at high R concentration ratios tend to be more hydrophilic than those for reduced GO. Furthermore, we see that the charges induced on the $sp^2$ carbon atoms, as considered by the CHELPG model, lead to a significant increase in the hydrophilicity of some species, particularly those with R = 10 and 60%. In the investigated cases, we can note that among all the concentration ratios, for only two of them (R = 20 and 30%) the CHELPG model presented values less favorable to hydration than the corresponding ones obtained with the average-charge model. It is interesting to note that although the charges at the $sp^2$ carbons tend to reduce the interaction energy, which corresponds to a reduction in the enthalpy, a uniform tendency is not observed in the entropic component since $\Delta G_{hyd}$ can increase or decrease depending on R. Finally, these results for free energy show that the charge redistribution on the carbon atoms can lead to significant changes in the hydrophilicity of graphene oxide. Such changes may become important in obtaining properties in the context of several applications, such as the use of GO as electrodes in supercapacitors or as inhibitors in processes involving biological molecules, where interaction with the environment plays a crucial role.

**Conclusions**

In this work we used atomistic molecular dynamics simulations to describe the hydration process of seven different models for graphene oxide at different levels of oxygenation. Two charge models for GO were considered: a simplified one, where $sp^2$ carbons were treated as LJ uncharged sites and another with CHELPG charges at all sites.

Mass density profiles and G(z) distributions of O··H pairs show that the structure of the hydration is practically unaffected by the models employed. Analysis of such profiles indicate that the water molecules near the surface of the GO are structured in two well defined layers. G(z) distribution of O··H pairs show that although the positions of the G(z) peaks are similar to those found in pure water, the height of the peaks is not, confirming the lower number of HBs formed between water and GO in relation to the number of HBs formed in bulk.

The interaction energy between water and GO correlates well with the number of hydrogen bonds formed, and both (HB's and energy) are significantly sensitive to the charge set employed. Our model employing CHELPG charges on all $sp^2$ carbons shows that the simplified model tends to overestimate the GO/water interaction energy.

The hydration free energy for each GO was determined as a function of the oxygen concentration. Our results show that for the investigated systems, $\Delta G_{hyd}$ values are in the range of -5 to -45 kJ mol$^{-1}$ indicating that hydration is a favorable process for all investigated systems. We found that the charge redistribution on the carbon atoms can lead to significant changes in the hydrophilicity of graphene oxide. This result is important since the interaction of GO with the medium may be fundamental for certain applications, such as the use of GOs as electrodes in supercapacitors or as inhibitors in processes involving biological molecules.

**Acknowledgments**

E.E.F. was supported by research grants from CAPES and FAPESP. V.V.C. did not obtain any funding.

**Contact Information**

E-mail: fileti@gmail.com. Fax: +55 12 3924 9500.

# REFERENCES

1. T. Tu, M. Lv, P. Xiu, T. Huynh, M. Zhang, M. Castelli, Z. Liu, Q. Huang, F. Chunhai, F. Haiping and R. Zhou, *Nat. Nanotech.*, 2013, 8, 594-601.
2. O. C. Compton, S. W. Cranford, K. W. Putz, Z. An, C. L. Brinson, M. J. Buehler and S. T. Nguyen, *ACS Nano*, 2012, 6, 2008-2019.
3. A. D. DeYoung, S.-W. Park, N. R. Dhumal, Y. Shim, Y. Jung and H. J. Kim, *The Journal of Physical Chemistry C*, 2014, 118, 18472-18480.
4. Q. Hu, B. Jiao, X. Shi, R. P. Valle, Y. Y. Zuo and G. Hu, *Nanoscale*, 2015, 7, 18025-18029.
5. L. Liu, L. Wang, J. Gao, J. Zhao, X. Gao and Z. Chen, *Carbon*, 2012, 50, 1690-1698.
6. K. P. Loh, Q. Bao, G. Eda and M. Chhowalla, *Nat. Chem.*, 2010, 2, 1015–1024.
7. N. V. Medhekar, A. Ramasubramaniam, R. S. Ruoff and V. B. Shenoy, *ACS Nano*, 2010, 4, 2300-2306.
8. Y. Cui, S. N. Kim, S. E. Jones, L. L. Wissler, R. R. Naik and M. C. McAlpine, *Nano. Lett.*, 2010, 4559–4565.
9. M. Zokaie and M. Foroutan, *RSC Advances*, 2015, 5, 97446-97457.
10. J. Zhang and D. Jiang, *Carbon*, 2014, 67, 784-791.
11. H. Tang, G. J. Ehlert, Y. Lin and H. A. Sodano, *Nano. Lett.*, 2012, 12, 84–90.
12. H. Lee, B. C. Ku and P. M. Ajayan, *Nano. Lett.*, 2012, 12, 1789–1793.
13. T. Si and E. T. Samulski, *Nano. Lett.*, 2008, 1679–1682.
14. A. F. Fonseca, H. Zhang and K. Cho, *Carbon*, 2015, 84, 365-374.
15. A. F. Fonseca, T. Liang, D. Zhang, K. Choudhary and S. B. Sinnott, *Computational Materials Science*, 2016, 114, 236-243.
16. S.-W. Park, A. D. DeYoung, N. R. Dhumal, Y. Shim, H. J. Kim and Y. Jung, *The Journal of Physical Chemistry Letters*, 2016, 7, 1180-1186.
17. X. Mu, X. Wu, T. Zhang, D. B. Go and T. Luo, *Scientific Reports*, 2014, 4, 3909.
18. N. Wei, C. Lv and Z. Xu, *Langmuir*, 2014, 30, 3572-3578.
19. N. Wei, X. Peng and Z. Xu, *Physical Review E*, 2014, 89, 12113.
20. J. Chen, X. Wang, C. Dai, S. Chen and Y. Tu, *Physica E: Low-dimensional Systems and Nanostructures*, 2014, 62, 59-63.
21. M. Feng, H. Kang, Z. Yang, B. Luan and R. Zhou, *The Journal of Chemical Physics*, 2016, 144, 225102.
22. A. M. Grant, H. Kim, T. L. Dupnock, K. Hu, Y. G. Yingling and V. V. Tsukruk, *Advanced Functional Materials*, 2016, 26, 6380-6392.
23. D. Stauffer, N. Dragneva, W. B. Floriano, R. C. Mawhinney, G. Fanchini, S. French and O. Rubel, *The Journal of Chemical Physics*, 2014, 141, 44705.
24. S. Zeng, L. Chen, Y. Wang and J. Chen, *Journal of Physics D: Applied Physics*, 2015, 48, 275402.
25. R. Devanathan, D. Chase-Woods, Y. Shin and D. W. Gotthold, *Scientific Reports*, 2016, 6, 29484.
26. J. Chen, G. Zhou, L. Chen, Y. Wang, X. Wang and S. Zeng, *The Journal of Physical Chemistry C*, 2016, 120, 6225-6231.
27. M. Zokaie and M. Foroutan, *RSC Advances*, 2015, 5, 39330-39341.
28. H. Tang, D. Liu, Y. Zhao, X. Yang, J. Lu and F. Cui, *The Journal of Physical Chemistry C*, 2015, 119, 26712-26718.
29. T. Dyer, N. Thamwattana and R. Jalili, *RSC Advances*, 2015, 5, 77062-77070.
30. C.-J. Shih, S. Lin, R. Sharma, M. S. Strano and D. Blankschtein, *Langmuir*, 2012, 28, 235-241.
31. C. M. Breneman and K. B. Wiberg, *Journal Of Computational Chemistry*, 1990, 11.
32. M. J. Frish and e. al., Gaussian, Inc., Wallingford CT, 2009.

33. R. B. Best, X. Zhu, J. Shim, P. E. Lopes, J. Mittal, M. Feig and A. D. Mackerell, Jr., *J Chem Theory Comput*, 2012, 8, 3257-3273.
34. W. L. Jorgensen, J. Chandrasekhar, J. D. Madura, R. W. Impey and M. L. Klein, *J. Chem. Phys.*, 1983, 79, 926-935.
35. T. Darden, D. York and L. Pedersen, *J. Chem. Phys.*, 1993, 98, 10089-10099.
36. G. Bussi, D. Donadio and M. Parrinello, *J. Chem. Phys. 126*, 2007, 126, 014101-014108.
37. M. Parrinello and A. Rahman, *J. Appl. Phys.*, 1981, 52, 7182-7192.
38. C. H. Bennett, *Journal of Computational Physics*, 1976, 22, 245-268.
39. M. R. Shirts, J. W. Pitera, W. C. Swope and V. S. Pande, *J. Chem. Phys.* , 2003, 119, 5740.
40. B. Hess, C. Kutzner, D. van der Spoel and E. Lindahl, *J. Chem. Theory Comput.*, 2008, 4, 435.
41. V. V. Chaban and E. E. Fileti, *New J. Chem.*, 2017, 41, 184-189
42. T. Malaspina, L. M. Abreu, T. L. Fonseca and E. Fileti, *Phys. Chem. Chem. Phys.*, 2014, 16, 17863-17868

**TOC**

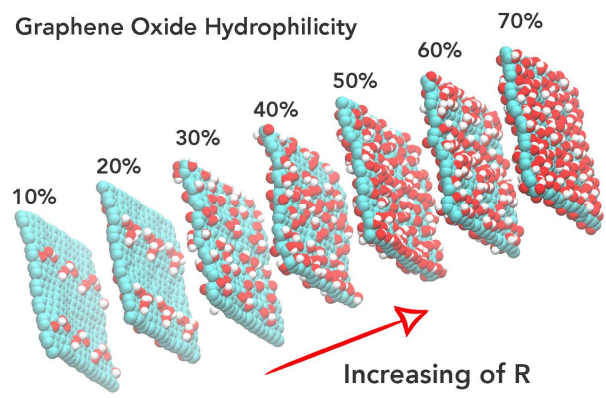